\documentclass[aps,prb,twocolumn,superscriptaddress, show pacs]{revtex4-2}

\usepackage{graphicx}
\usepackage{dcolumn}
\usepackage{bm}
\usepackage{hyperref}
\usepackage{bm, amsmath}
\usepackage{txfonts}

\begin{document}

\title{Confinement-Induced Metastability and Structural Diversity of Hopfions in Chiral Magnetic Films}

\author{Andrey O. Leonov}
\thanks{Corresponding author: leonov@hiroshima-u.ac.jp}
\affiliation{Department of Chemistry, Faculty of Science, Hiroshima University Kagamiyama, Higashi Hiroshima, Hiroshima 739-8526, Japan}
\affiliation{International Institute for Sustainability with Knotted Chiral Meta Matter, Kagamiyama, Higashi Hiroshima, Hiroshima 739-8526, Japan} 

\author{Takayuki Shigenaga}
\affiliation{Department of Chemistry, Faculty of Science, Hiroshima University Kagamiyama, Higashi Hiroshima, Hiroshima 739-8526, Japan}
\affiliation{International Institute for Sustainability with Knotted Chiral Meta Matter, Kagamiyama, Higashi Hiroshima, Hiroshima 739-8526, Japan} 

\date{\today}

\begin{abstract}
Topological particle-like excitations such as skyrmions and hopfions offer rich opportunities for spintronic and photonic applications. While skyrmions have been extensively studied, the stabilization mechanisms and phase behavior of three-dimensional hopfions remain largely unexplored. Here, we investigate the formation, stability, and interactions of hopfions in thin chiral magnetic films with surface anchoring, using three-dimensional micromagnetic simulations within a material-independent framework applicable to both magnetic and liquid crystalline systems. We identify four distinct types of isolated hopfions, generated by rotating bimeron and finger-like solitons around a central axis. The metastability regions of these precursor textures closely follow the boundaries of modulated finger phases, enabling their size to be continuously tuned through anisotropy-driven inflation and collapse. Remarkably, we demonstrate that hopfions near their inflation threshold possess energies comparable with the homogeneous state, allowing them to enclose regions of modulated phases or other solitons, forming higher-order, bag-like domains. In contrast, periodic hopfion lattices remain intrinsically unstable under confinement, spontaneously relaxing into finger phases. These findings establish general principles for stabilizing, tuning, and assembling three-dimensional topological solitons in confined chiral systems, suggesting experimentally accessible routes for texture engineering in liquid crystals via electric-field control.
\end{abstract}

\maketitle

\section{Introduction.} 

The concept of topological solitons transcends disciplinary boundaries, arising in effective field theories describing systems from nuclear to condensed matter physics~\cite{manton_sutcliffe,shnir}. A soliton is a self-stabilizing, particle-like state in a continuous field~\cite{solitons,Volovik}, existing as a finite, localized excitation embedded within a uniform vacuum background, and maintaining its integrity through topological or dynamical conservation laws. Beyond isolated objects, solitons can self-organize through their mutual interactions into periodic arrays or crystal-like assemblies. This gives rise to a form of \emph{solitonic meta-matter}, in which individual solitons function as quasi-atoms of an emergent material.

\begin{figure*}[t]
  \centering
  \includegraphics[width=0.8\linewidth]{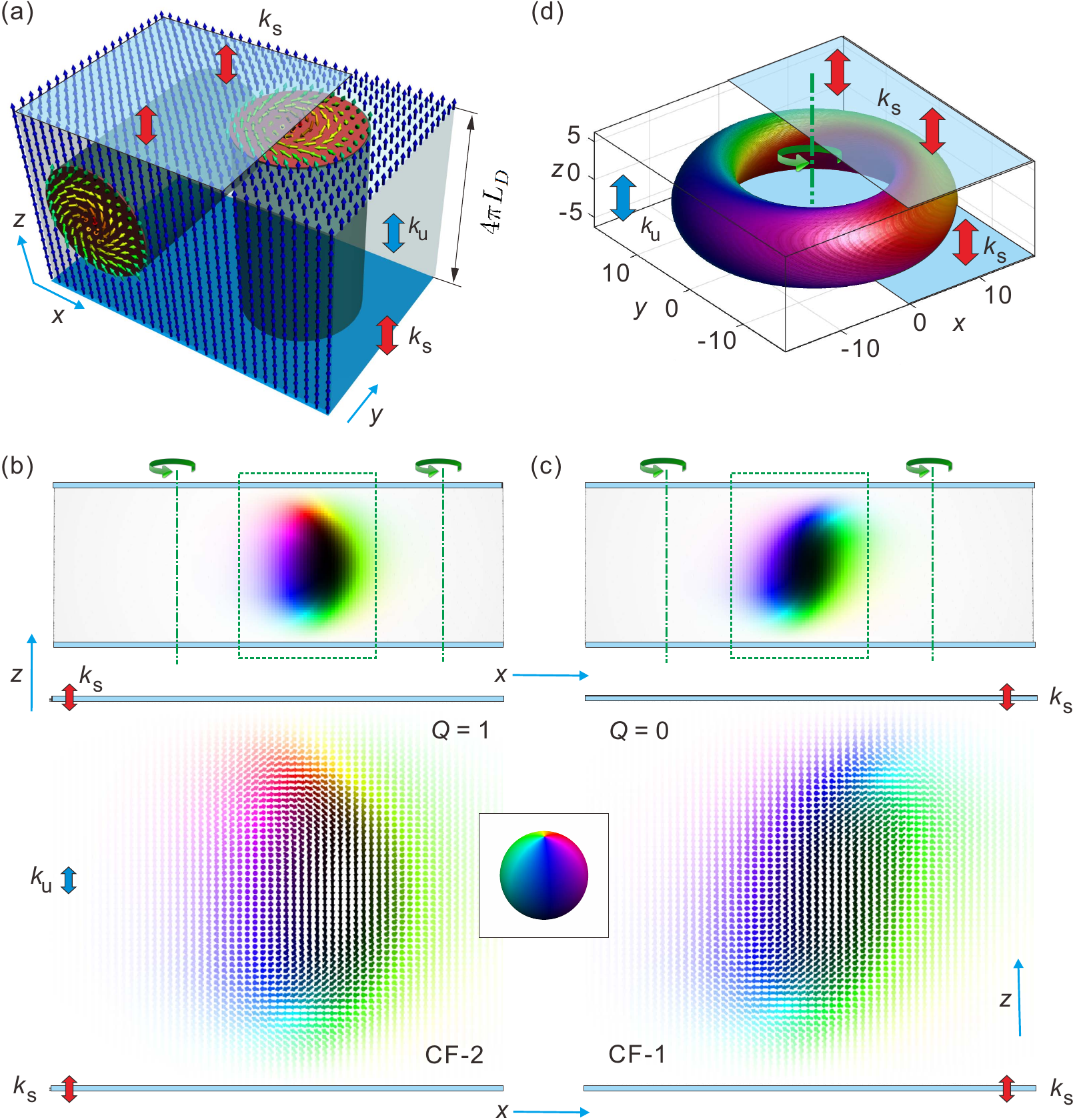}
  \caption{\label{fig01} \textrm{Creation of hopfions by rotating CF-1 and CF-2 fingers.}
  (a) Schematics of two skyrmion filaments: one aligned with the surface normal and another oriented perpendicular to it. To accommodate the surrounding homogeneous state, the axisymmetric skyrmion structure must continuously deform into the configuration shown in (b).
  (b,c) Spin distributions within the two-dimensional CF-1 and CF-2 solitons. The green dash-dotted lines mark the rotation axes used to generate the corresponding three-dimensional hopfion textures.
  (d) A representative 3D hopfion obtained through this rotational construction.}
\end{figure*}

Among topological solitons, two-dimensional (2D) particle-like skyrmions~\cite{Bogdanov94} have attracted significant attention in condensed matter systems~\cite{NT}. Skyrmions are classified by the second homotopy group~\cite{Faddeev,Bott}, $\pi_2(S^2) = \mathbb{Z}$, and possess an integer-valued topological charge, $Q = \frac{1}{4\pi} \int \mathbf{m} \cdot \left( \frac{\partial \mathbf{m}}{\partial x} \times \frac{\partial \mathbf{m}}{\partial y} \right) \, dx\, dy$,
where $\mathbf{m}$ is a continuous unit vector field. Skyrmions have been widely investigated in both chiral magnets (ChM)~\cite{Fert2017,Back2020} and chiral liquid crystals (CLC)~\cite{Oswald} due to their ability to exhibit novel physical phenomena and their potential for future spintronic and photonic applications~\cite{Fert2013, smalyukh_review}. 

In bulk chiral magnets, e.g., in MnSi~\cite{Muhlbauer} and FeGe~\cite{YuFeGe}, axisymmetric skyrmionic filaments typically align parallel to an applied magnetic field or the easy axis of magnetic anisotropy (Fig.~\ref{fig01}(a)). When their orientation switches perpendicular to this axis, a skyrmion transforms into a pair of coupled merons, known as a bimeron (or a cholesteric finger CF-2 in CLC~\cite{Oswald}), each carrying a fractional topological charge $Q = 1/2$, thereby preserving its overall topological identity while matching the surrounding homogeneous state~\cite{toggle,duzgun2018}.

A hopfion can be understood as a three-dimensional (3D) extension of this idea~\cite{magnetism,metlov2025}: a 2D bimeron-like structure continuously rotated around an axis in three dimensions, forming a toroidal, topologically non-trivial soliton characterized by a Hopf invariant~\cite{Whitehead,Gladikowski}, $Q_H = \frac{1}{(4\pi)^2} \int \mathbf{A} \cdot \left( \nabla \times \mathbf{A} \right) \, d^3 r$, where $\mathbf{A}$ is a vector potential for the emergent magnetic (or helicity) field $\nabla \times \mathbf{A}$ associated with the hopfion~\cite{guslienko2024}. These configurations are classified by the third homotopy group, $\pi_3(S^2) = \mathbb{Z}$~\cite{sutcliffe_review}. Unlike skyrmions and bimerons, hopfions are inherently 3D, with distinct topological, energetic, and dynamical properties~\cite{kent_2021,saji2023,Gobel2020,Khodzhaev2022} (Fig.~\ref{fig01}(d)).

Although the concept of hopfions has been theoretically proposed in various fields, their realization has been predominantly explored in confined geometries, such as thin magnetic~\cite{YuHopfion} or liquid crystalline films~\cite{ackerman_2017}. Confinement plays a pivotal role in stabilizing hopfions. 
For instance, in bulk chiral magnets, hopfions tend to elongate along the direction of the applied magnetic field, eventually relaxing into spiral phases~\cite{magnetism,metlov2025}. In contrast, thin film geometries impose finite-size effects and boundary conditions that inhibit such elongations and dramatically reshape the topological phase landscape. Notably, the conical spiral phase, typically dominant in bulk chiral magnets (the reason why, e.g., skyrmions were found only in small A-phase pockets near the Curie temperatures in chiral magnets~\cite{Muhlbauer}), loses rotational energy near surfaces where magnetization is pinned by surface anchoring~\cite{anchoring}, while bimeron-based chain configurations and their derivatives become energetically favorable. This reshaping of the phase diagram not only broadens the stability window for bimeron-based hopfions but also permits the emergence of novel 3D textures based on CF-1 finger-like structures with $Q = 0$~\cite{Oswald}, which have no counterparts in bulk magnets (Fig.~\ref{fig01}(c)).

In this work, we systematically investigate the stabilization, structure, and interactions of hopfions in thin chiral magnetic films with surface anchoring. Through 3D micromagnetic simulations, we identify four distinct types of isolated hopfions, classified by their internal structure: two formed by continuous rotation of bimerons and two derived from the rotation of CF-1 textures (Fig.\ref{fig02}). These configurations differ by the position of the rotation axis relative to the corresponding two-dimensional soliton, either to its left or right (green dash-dotted lines in Fig.\ref{fig01}(b),(c)).

Isolated hopfions are found to exist in close proximity to the phase boundaries of the corresponding finger phases, revealing the essential role of confinement-induced topological frustration. By mapping the phase diagram in the plane of uniaxial anisotropy and surface anchoring strength—two parameters directly tunable in both magnetic~\cite{Yamamoto,Johnson1996} and liquid crystalline systems~\cite{Oswald}—we demonstrate that varying the uniaxial anisotropy, equivalent to applying an electric field in chiral liquid crystals~\cite{Oswald}, induces qualitative transformations in the behavior of isolated hopfions: either gradual inflation and deformation into finger textures as the phase boundary is approached, or shrinkage and collapse into toron-like structures under increased anisotropy. Furthermore, we reveal that the distinct radii of different hopfion types (as the phase boundaries of CF-1 and CF-2 phases do not coincide) enable the formation of composite configurations, where a smaller hopfion can be encapsulated within a larger one, with their sizes mutually adjusted to minimize repulsive interactions. In particular, the external hopfion can substantially expand its radius to alleviate overlap with the internal soliton, offering a mechanism for engineering multi-soliton states. More generally, the interior of a topological bag  can be filled by a group of torons or the domain of a finger phase. 

Finally, we demonstrate that periodic hopfion lattices (HL) cannot be stabilized under these conditions, as their energy landscape lacks a local minimum, and readily relaxes into the modulated finger phase. These findings introduce new mechanisms for stabilizing, tuning, and combining three-dimensional topological solitons in confined chiral systems, providing practical avenues for topological texture engineering in both magnetic and liquid crystalline materials.

\begin{figure}[t]
    \centering
    \includegraphics[width=0.49\textwidth]{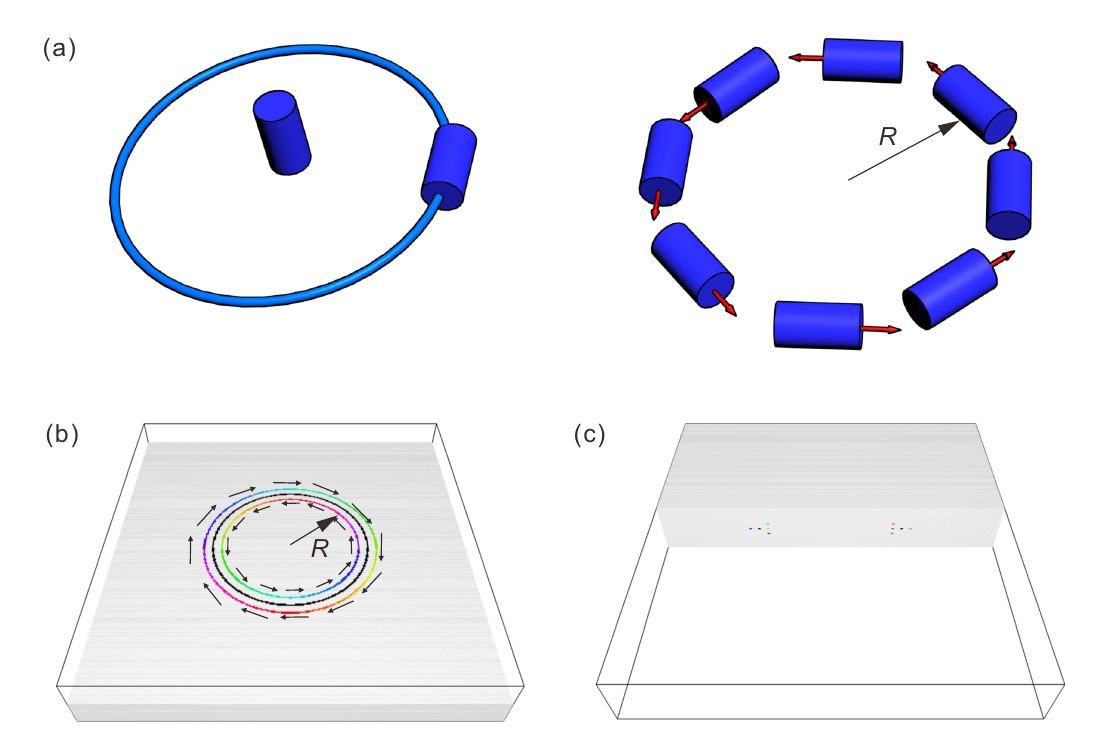}
    \caption{\textrm{Procedure for constructing initial hopfion configurations for relaxation in \textsc{mumax3}.} 
    (a) Elementary cylinders are positioned along prescribed circular trajectories and assigned magnetization vectors of specified orientation. 
    (b) Positions of circular trajectories in the middle plane of the film. 
    (c) Five circular trajectories visualized in $xz$-plane cross-sections.}
    \label{fig01v}
\end{figure}

\begin{figure*}[t]
  \centering
  \includegraphics[width=0.8\linewidth]{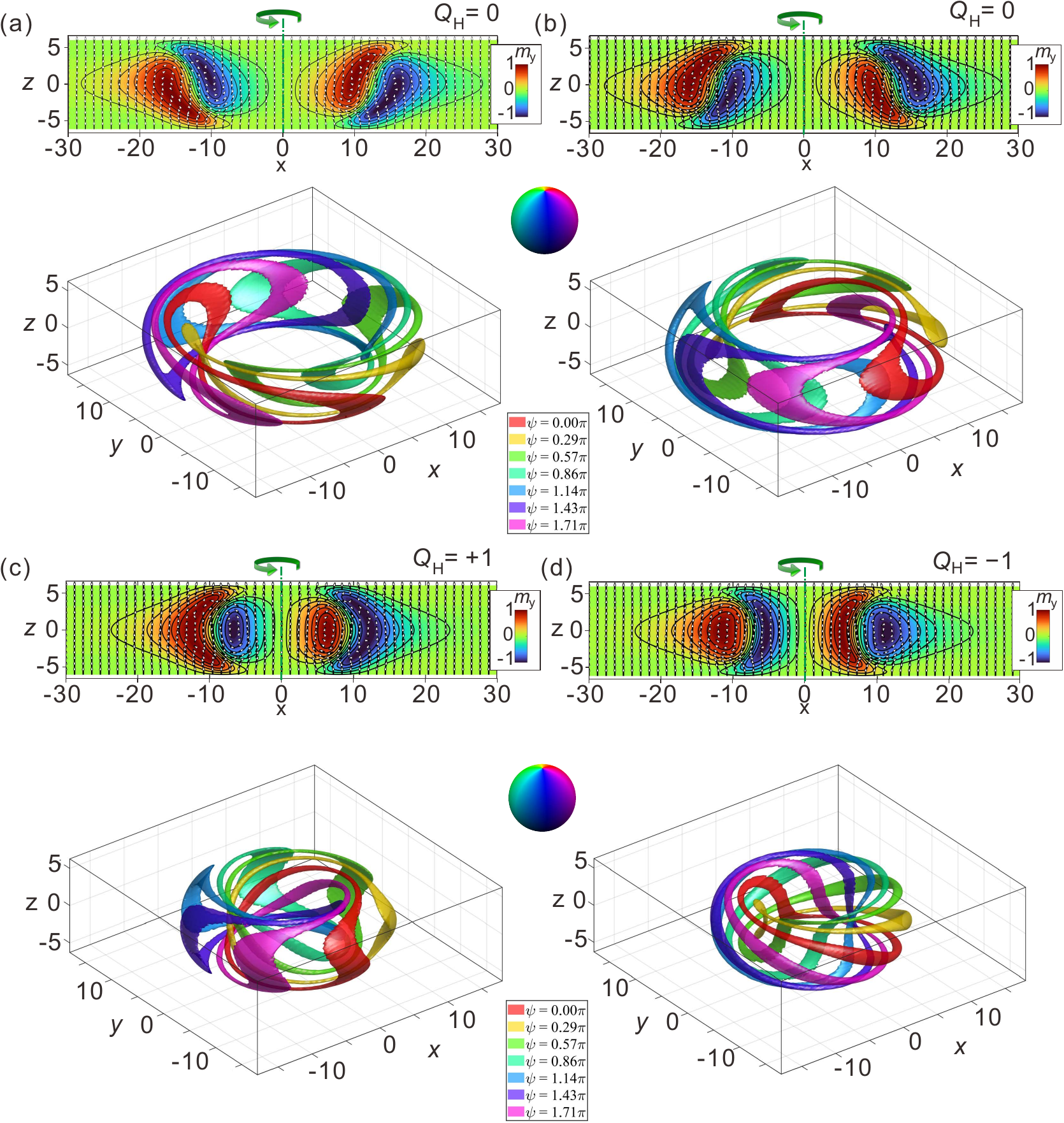}
  \caption{\label{fig02} \textrm{Four types of isolated hopfions.} (a, b) Trivial hopfions with Hopf index $Q_H = 0$. (c, d) Hopfions with $Q_H = +1$ and $Q_H = -1$, respectively. Upper panels show color maps of the $m_y$ component in $xz$ cross-sections. Lower panels display preimages plotted for $\theta = \pi/2$ and several values of the azimuthal angle $\psi$.}
\end{figure*}

\section{The phenomenological model.}

The magnetic energy density of a noncentrosymmetric ferromagnet that contains both bulk ($\Omega$) and surface ($\partial \Omega$) terms is expressed as the sum of exchange, Dzyaloshinskii–Moriya interaction (DMI), and uniaxial anisotropy contributions~\cite{anchoring}:
\begin{align}
W(\mathbf{m}) = &\int_{\Omega} \left( (\nabla \mathbf{m})^2 + \mathbf{m} \cdot (\nabla \times \mathbf{m}) - k_u (\mathbf{m} \cdot \mathbf{z})^2 \right) d^3 \mathbf{r}
\nonumber\\
&-\int_{\partial \Omega} k_s (\mathbf{m} \cdot \mathbf{z})^2 \, d^2 \mathbf{r}.
\label{functional}
\end{align}
Here, we employ dimensionless units to generalize the results and allow direct mapping to various material systems. The characteristic length scale is $L_D = A/D$, and the dimensionless anisotropy constant is defined as $k_u = K_u A / D^2$.
The system geometry corresponds to a thin film, infinite in the $x$- and $y$-directions (i.e., periodic boundary conditions are imposed along these directions), with a thickness along $z$ equal to a single spiral pitch, $4\pi L_D$. 
Such a geometry is analogous to trilayered ChM systems, consisting of a single chiral ferromagnetic layer with weaker perpendicular magnetic anisotropy (PMA) sandwiched between two capping ferromagnetic layers with stronger PMA \cite{Hu2025}.

Both $k_u$ and the surface anchoring strength $k_s$ promote an easy-axis ferromagnetic state with magnetization aligned along $\mathbf{z}$.

Importantly, the theoretical frameworks governing ChM and CLC are formally equivalent within the widely used one-constant approximation for liquid crystals ~\cite{smalyukh_review,anchoring}. In this approximation, the energy functionals describing both systems share the same mathematical form, differing only in the physical interpretation of the order parameter: the magnetization vector $\mathbf{m}$ in chiral magnets and the director field $\mathbf{n}$ in liquid crystals~\cite{Oswald, smalyukh_review}. Consequently, theoretical predictions made for one system can be directly transferred to the other, provided appropriate material parameters and boundary conditions are considered. 
In addition, CLC offer several advantages over magnetic systems for the investigation of inhomogeneous structures. Their system parameters can be tuned over a broad range to establish the desired experimental conditions; experiments are typically conducted at room temperature and are relatively straightforward; and, most importantly, the resulting structures are readily visualized—often with a degree of clarity that is difficult to achieve in magnetic systems. Nevertheless, in ferromagnetic materials, advanced techniques such as electron holography and X-ray magnetic imaging have recently enabled the experimental visualization of complex three-dimensional magnetization configurations~\cite{donnelly2020}.
This equivalence has enabled valuable cross-disciplinary insights, with liquid crystals serving as a versatile experimental platform for realizing and visualizing topological solitons originally proposed in magnetic systems, and vice versa~\cite{smalyukh_review,ackerman_2017}.
The uniaxial anisotropy $k_u$ in CLC is typically imposed by an applied electric field $\mathbf{E}$, contributing an energy term of the same functional form, $-(\varepsilon_0 \Delta \varepsilon / 2) (\mathbf{n} \cdot \mathbf{E})^2$, where $\Delta \varepsilon$ is the dielectric anisotropy. 
The Zeeman energy term, $-\mathbf{m} \cdot \mathbf{h}$, is omitted throughout this work, as it has no direct analog in liquid crystal systems. In addition, such a linear coupling does not exist in standard apolar nematic liquid crystals, further justifying its exclusion.

In this work we adopt a unified terminology motivated by both chiral magnetism and chiral liquid-crystal research. For instance, following established practice in chiral magnets, we refer to CF-2 cholesteric fingers as \emph{bimerons}, since their cross-sectional structure, topological charge distribution, and characteristic double-core morphology are fully analogous to magnetic bimerons. In contrast, there is no direct magnetic analogue or widely accepted term for CF-1 structures. Our goal is to bridge the vocabulary and conceptual frameworks used in chiral magnetism and chiral liquid crystals to facilitate a more coherent dialogue between the two communities.

\begin{figure*}[t]
    \centering
    \includegraphics[width=1\textwidth]{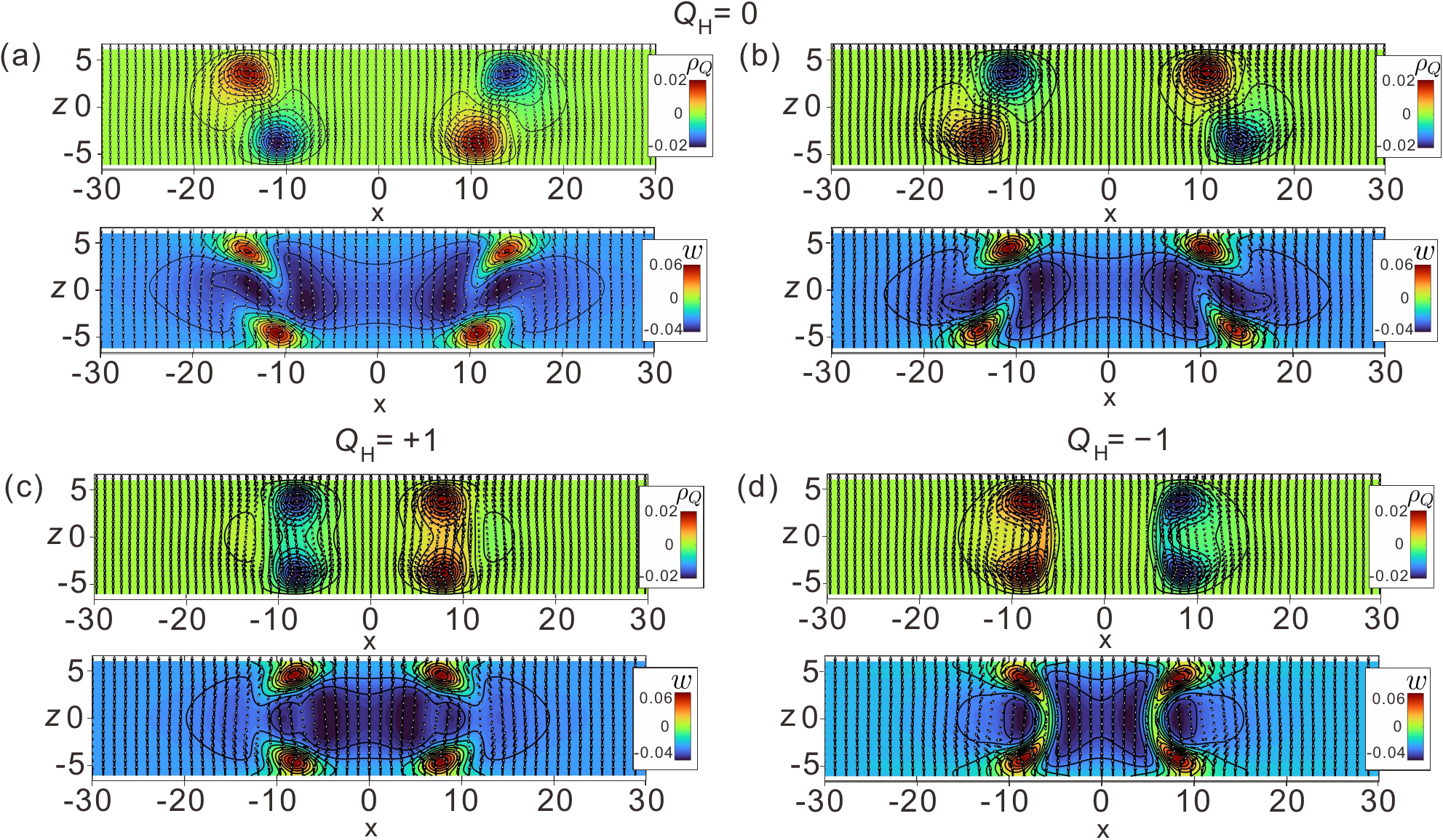}
    \caption{\textrm{Topological charge density $\rho_Q$ and local energy density $w$ for isolated hopfions.} Color maps show the distributions of $\rho_Q$ and $w$ in the $xz$-plane for all hopfion varieties at $k_u = 0.22$, $k_s = 50$.}
    \label{fig:A1}
\end{figure*}

\section{Isolated hopfions.}

\subsection{Initial Hopfion Configurations\label{A1}}

Initial hopfion configurations were created in \textsc{mumax3} \cite{mumax3} by arranging the magnetization vectors along several circular trajectories. First, a small cylinder was defined at the center of the numerical grid (Fig.~\ref{fig01v}(a)), which was rotated by $\pi/2$ around the $x$-axis and then by an angle $\alpha$ around the $z$-axis. The angle $\alpha$ was chosen such that the cylinder axis became tangent to a predefined circular trajectory with a fixed radius.

These elementary cylinders were then translated to align seamlessly along the corresponding circular path. The magnetization direction within each elementary cylinder was set so that, along the circle of radius $R$, the magnetization vector was co-aligned with the local cylinder axis (Fig.~\ref{fig01v}(b)). The value of $R$ corresponds to the hopfion radius consistently reported in the main text for all hopfion varieties.

Typically, several such circular trajectories (Fig.~\ref{fig01v}(c) shows five trajectories for a $Q_H=1$ hopfion) with prescribed and pinned magnetization orientations were sufficient to generate a smooth hopfion structure after energy minimization in \textsc{mumax3} \cite{mumax3}. For example, in Fig.~\ref{fig01v}(b), the magnetization is aligned counterclockwise on the inner circle, points downward on the middle circle, and rotates clockwise at the outermost trajectory.

The simulation grid size was set to $320 \times 320 \times 64$, with a cell size along the $z$-direction of $4\pi/64$.

This procedure for creating initial hopfion states does not rely on any Ansatz solutions used, e.g., in Ref. \cite{Hu2025} and can be readily adapted to construct hopfions of arbitrary complexity.

To obtain the energy dependencies presented later  in Fig.~\ref{fig04}(a),(b), only the magnetization along the circular path with radius $R$ was pinned while computing the eigen hopfion energy $W(R)$.

\begin{figure}[t]
    \centering
    \includegraphics[width=0.4\textwidth]{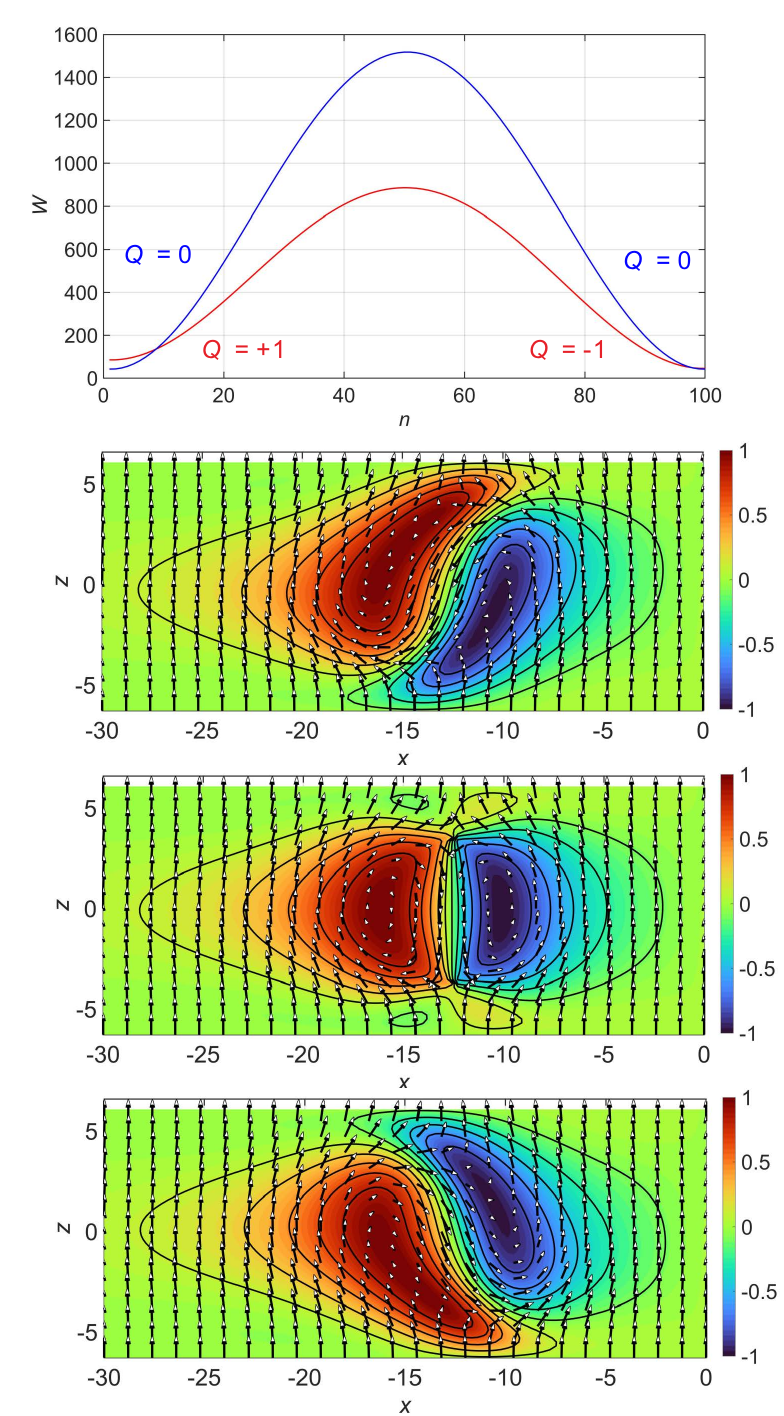}
    \caption{\textrm{Energy barriers between different hopfion states.}
    The blue curve corresponds to $Q_H = 0$ hopfions, while the red curve represents $Q_H = \pm 1$ hopfions. Color plots depict three characteristic spin configurations for trivial $Q_H = 0$ hopfions along the transition pathway.}
    \label{fig03v}
\end{figure}

\subsection{Internal structure of isolated hopfions}

To elucidate the internal structure of the four types of isolated hopfions identified in our simulations, we computed and visualized several key field characteristics in the $xz$-plane, including the magnetization distribution (upper panel in Fig.~\ref{fig02}), the topological charge density $\rho_Q$, and the local energy density $w$ (Fig.~\ref{fig:A1}). Additionally, we reconstructed preimage surfaces for selected magnetization directions (with $\theta = \pi/2$ and varying azimuthal angle $\psi$) to reveal the global topology of these solitons (lower panel in Fig.~\ref{fig02}).

Remarkably, while the hopfions based on CF-1 fingers are energetically degenerate and possess a zero Hopf index, $Q_H = 0$ (Fig.~\ref{fig02}(a),(b)), the two hopfions constructed from CF-2 fingers (Fig.~\ref{fig02}(c),(d)) represent genuinely distinct topological solitons with $Q_H = \pm 1$. 

Note that we also refer to the solitons generated by rotating the CF-1 fingers as hopfions, even though their Hopf index formally vanishes. This terminology is consistent with Ref.~\cite{Ackerman2017}, where the term “hopfion’’ is applied to a broader class of three-dimensional, torus-like solitons characterized by a closed-loop preimage structure. The CF-1–derived textures preserve the essential geometrical features associated with hopfions—namely, a twisted toroidal core and mutually linked isosurfaces—despite having $Q_{\mathrm{H}} = 0$. Thus, their classification as hopfions emphasizes their morphological and dynamical similarity to genuine Hopf-charge-carrying configurations, and aligns our nomenclature with the emerging literature in chiral magnetic and liquid-crystal systems.

Experimentally, different types of hopfionic textures are well distinguished by their characteristic geometric and optical signatures. In particular, the three-dimensional director-field configurations associated with CF-1-- and CF-2--derived solitons produce distinct patterns under polarizing optical microscopy and three-photon excitation fluorescence imaging. As demonstrated in Ref.~\cite{Ackerman2017}, variations in the internal twist, core morphology, and linking structure of preimages lead to clearly identifiable experimental fingerprints. This ensures that, despite differences in their topological indices or origins, all hopfion types considered here can be reliably differentiated in practice.

\subsection{Transformation among hopfions \label{A2}}


In the present section, we investigate possible pathways for transformations between different hopfion configurations, such as between two trivial hopfions with $Q_H=0$ and between hopfions with $Q_H=\pm 1$ (see  Supplementary Videos~1–4 for further details and Ref.~\cite{bessarab2015}). As shown in Fig.~\ref{fig03v}, hopfion states are separated by significant energy barriers, which  are much larger than the energies of the metastable hopfions themselves. Color plots display the spin configurations in the $xz$ plane for trivial hopfions corresponding to the energy minima and the saddle-point configuration (middle panel). The interpolation between hopfions proceeds via the formation of localized defects. In this respect, the spin texture cross-section at the saddle point resembles the structure of a toron.



\begin{figure*}[t]
  \centering
  \includegraphics[width=0.8\linewidth]{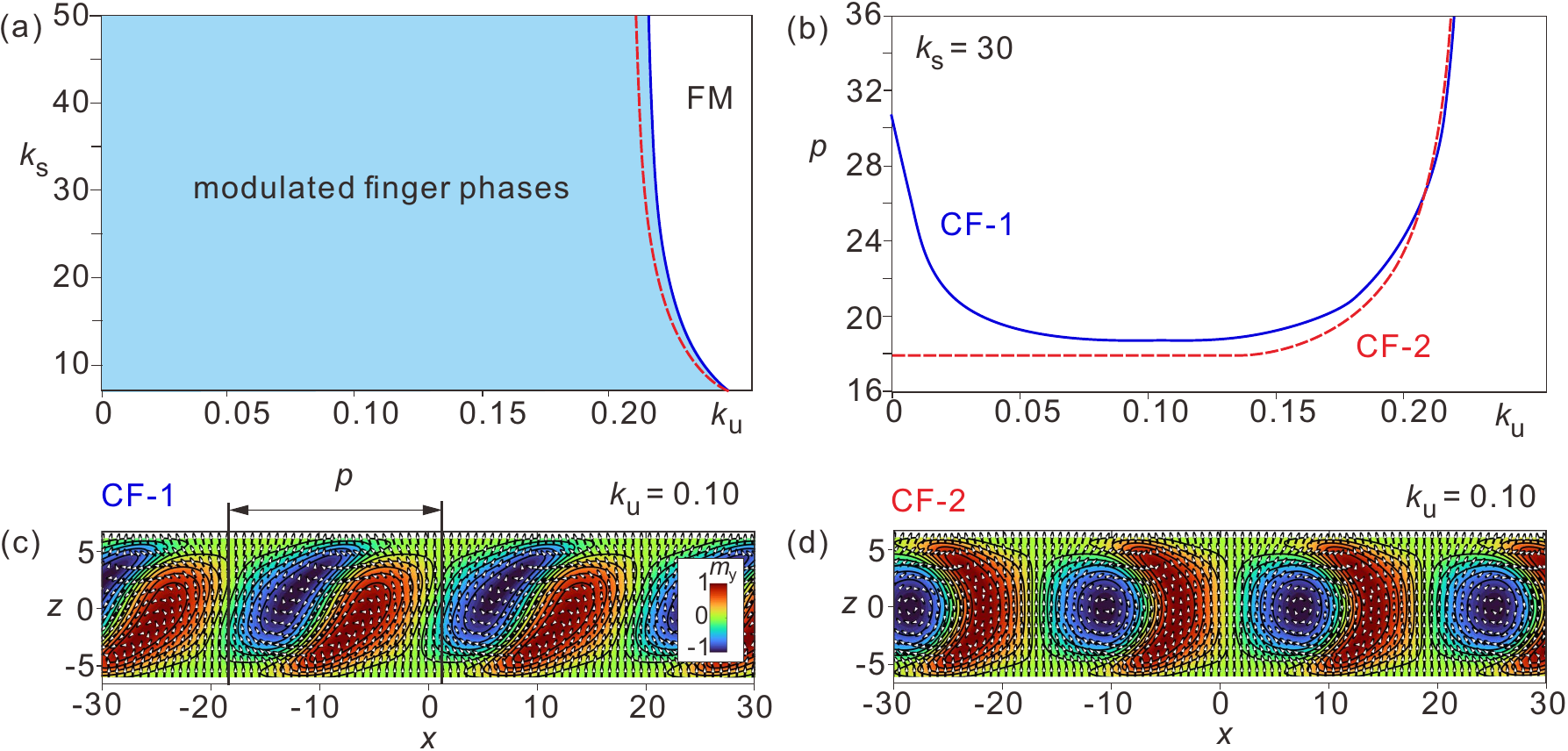}
  \caption{\label{fig03} \textrm{Phase diagram and internal structure of modulated finger phases.}  (a) Phase diagram in the space of control parameters $(k_u, k_s)$. To the left of the solid blue (dashed red) line, the energy of a periodic CF-1 (CF-2) phase becomes lower than that of the homogeneous state. (b) Equilibrium periods of the finger phases as a function of $k_u$ for a fixed surface anchoring $k_s = 30$, illustrating their expansion and the release of isolated fingers. (c, d) Magnetization distributions within two-dimensional CF-1 and CF-2 phases, respectively.}
\end{figure*}


To construct initial transition paths between two metastable hopfion configurations, we employed a direct geodesic interpolation method in spin space. The spin configurations $\{\mathbf{m}^A_i\}$ and $\{\mathbf{m}^B_i\}$ at each lattice site $i$ were obtained from relaxed micromagnetic simulations using \textsc{mumax3}. For each lattice site, the intermediate magnetization vector at interpolation parameter $t \in [0,1]$ was computed via spherical linear interpolation (slerp) according to
\begin{equation}
    \mathbf{m}_i(t) = \frac{\sin[(1-t)\theta_i]}{\sin \theta_i} \, \mathbf{m}^A_i
                      + \frac{\sin(t\theta_i)}{\sin \theta_i} \, \mathbf{m}^B_i,
\end{equation}
where $\theta_i = \arccos(\mathbf{m}^A_i \cdot \mathbf{m}^B_i)$ is the angle between the corresponding spins in the two configurations. This interpolation preserves the unit length of the magnetization vectors at each site.

The total number of intermediate images generated along the path was typically $N=100$. Each interpolated configuration was then optionally relaxed using overdamped Landau-Lifshitz-Gilbert dynamics (with damping constant $\alpha=1$) in an attempt to remove sharp local defects without allowing the system to fully collapse into a local energy minimum. The effective magnetic field used during the relaxation included contributions from dimensionless exchange, bulk DMI, uniaxial anisotropy, and surface anisotropy terms, consistent with the micromagnetic model employed in \textsc{mumax3}.

This approach provided a continuous set of physically reasonable intermediate states suitable for subsequent energy profile evaluation along the transition path.

\section{Modulated finger phases.}

\subsection{Anisotropy-driven expansion of modulated finger phases}

Since the internal structure of isolated hopfions is directly related to the structure of the underlying finger textures, it is instructive to investigate the stability regions of these finger phases alongside their anisotropy-driven evolution. In the broad region of control parameters $k_s$ and $k_u$ (Fig.~\ref{fig03}(a)), the modulated CF-1 phase with an equilibrium period $p$ represents the local energy minimum of the functional~(\ref{functional}). This stability is particularly pronounced for large values of the surface anchoring strength, typical for chiral liquid crystals (blue-shaded region in Fig.~\ref{fig03}(a). 

As the uniaxial anisotropy $k_u$ increases, 
the period of the CF-1 phase initially decreases (blue solid line in Fig.~\ref{fig03}(b)) and subsequently diverges at a critical value of $k_u$ (e.g., $k_u \approx 0.216$ for $k_s = 50$). This behavior closely resembles the magnetic field-driven evolution of spiral and skyrmion states in chiral magnets~\cite{Togawa12,Kovacs2017}. Beyond this critical anisotropy, isolated fingers become metastable solitons with positive energy relative to the homogeneous state.

Bimerons exhibit a qualitatively similar evolution, although the critical value of $k_u$ at which isolated bimerons are released is slightly lower (dashed red line in Fig.~\ref{fig03}(b); for instance, $k_u \approx 0.212$ at $k_s = 50$). 

Representative magnetization distributions of finger textures are presented in Fig.~\ref{fig03}(c),(d).

\subsection{Phase Diagram of States \label{A3}}

\begin{figure}[t]
    \centering
    \includegraphics[width=0.48\textwidth]{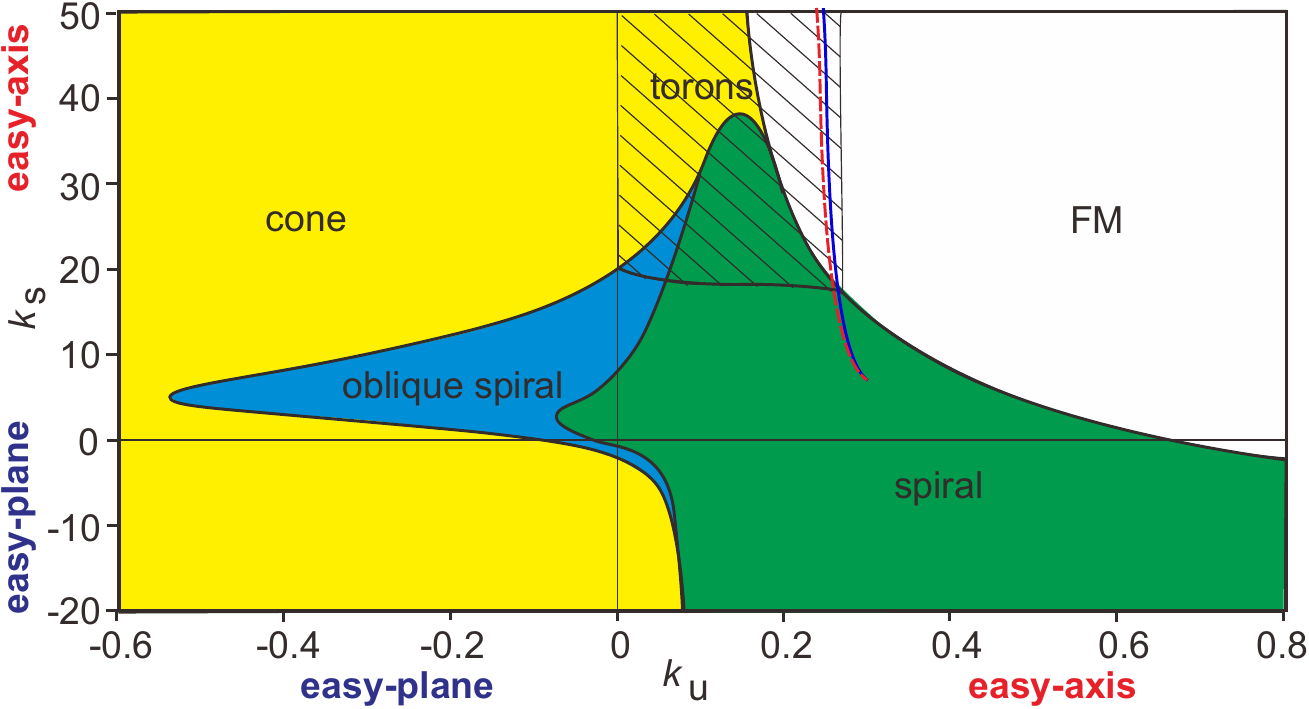}
    \caption{\textrm{Phase diagram of states.}
    The regions of thermodynamical stability for different phases are shown in distinct colors. A detailed description of the phase transitions between the modulated phases is provided in Ref.~\cite{anchoring}.}
    \label{fig04v}
\end{figure}

We also note that other phases present in the system can be safely omitted from the present analysis, although toron lattices and spiral states represent lower-energy configurations (Fig. \ref{fig04v}). This omission is justified by the topological independence of each modulated phase: geometrical and topological constraints prevent continuous transformations between them under confinement. 

The phase diagram of states in the space of control parameters $(k_u, k_s)$ is presented in  Fig.~\ref{fig04v} for both signs of the surface and bulk anisotropies \cite{anchoring}. The regions of thermodynamic stability for different phases are indicated by distinct colors. The hatched region marks the domain where the toron lattice represents the global energy minimum. In the present work, we focus exclusively on the upper-right quadrant of this phase diagram.

Notably, although the finger phases are overshadowed by toron lattices at large $k_s$ values and by spiral phases at small $k_s$, there exists no continuous topological path allowing their transformation into these more energetically favorable states. This topological isolation stabilizes the finger phases as metastable textures within their respective regions of the phase diagram.

\begin{figure*}[t]
  \centering
  \includegraphics[width=0.8\linewidth]{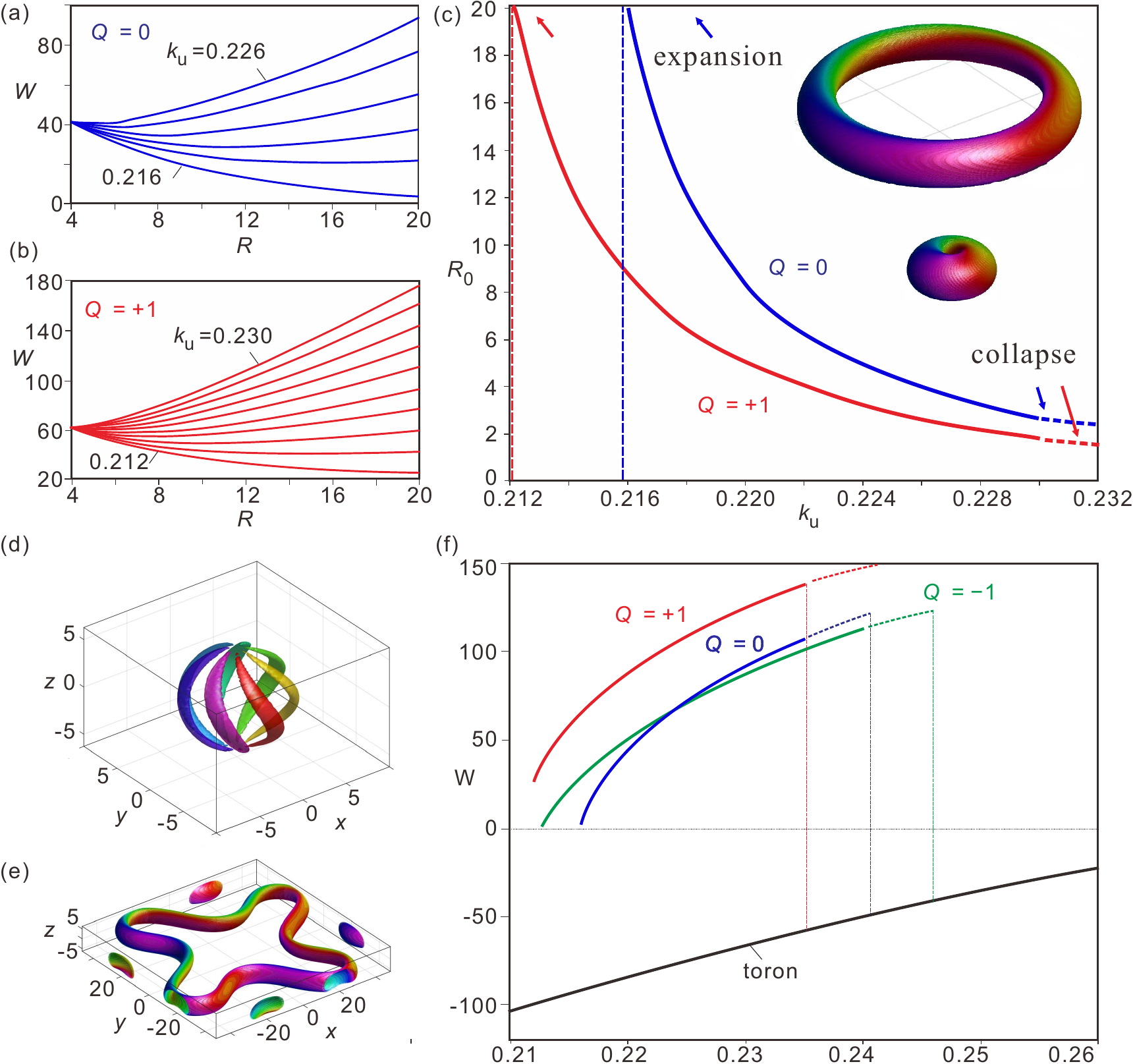}
  \caption{\label{fig04} \textrm{Energetic and structural characteristics of isolated hopfions.}  (a, b) Eigen-energies of isolated hopfions computed relative to the FM state and plotted as a function of the pinned hopfion radius $R$ for various values of $k_u$. For illustrative purposes, all curves except the bottom one are vertically shifted by individual energy offsets $\Delta W$ such that their values coincide at $R = 4$. (c) Equilibrium radii $R_0$ as a function of $k_u$, illustrating the inflation of hopfions near the phase boundary and their collapse into torons at large $k_u$ values. 
(d) Preimage structure corresponding to a toron. 
(e) Transformation of an isolated hopfion into a modulated finger phase once $k_u$ exceeds the metastability range.  (f) Eigen-energies of hopfions as a function of $k_u$.}
\end{figure*}

We note that the phase diagram presented in Ref.~\cite{Hu2025} for trilayered systems omits several of the phases discussed here.

\subsection{Isolated hopfions: from collapse to expansion.}

The critical lines of the finger phases in Fig.~\ref{fig03}(a) serve as reliable guides for identifying the stability regions of metastable hopfions. The eigen-energies $W$ of isolated hopfions, 
plotted in Fig.~\ref{fig04}(a),(b) for a fixed surface anchoring strength $k_s$ and varying uniaxial anisotropy $k_u$, exhibit distinct minima at equilibrium radii $R_0$ located in the immediate vicinity of the corresponding finger phase boundaries. This confirms that metastable hopfions act as precursor textures stabilized by proximity to modulated finger phases. Notably, the equilibrium radii differ between hopfion types, providing a practical criterion for their experimental identification based on size.

As shown by the dependence of $R_0(k_u)$ in Fig.~\ref{fig04}(c), all hopfion types collapse into toron-like structures with increasing electric field 
(Fig.~\ref{fig04}(d)). In this regime, the specific identity of the initial hopfion type is effectively erased, as all configurations converge to a common toron topology. Conversely, as the system approaches the stability regions of the CF-1 and CF-2 phases, hopfions inflate without bound.

When the electric field is reduced beyond the metastability window, isolated hopfions become elliptically unstable and begin to elongate in order to occupy the available space, thereby reconstructing the corresponding modulated finger phase whose equilibrium period was calculated in Fig.~\ref{fig03}. An example of such a distorted hopfion is shown in Fig.~\ref{fig04}(e). At the initial stage of this deformation, the hopfion does not immediately lose its identity: it still exhibits linked preimages, indicating that its core topological structure remains intact. However, as $k_u$ decreases further, the equilibrium period of the finger phase shrinks, requiring increasingly dense packing of fingers. This increasing geometric frustration is expected to eventually induce ruptures in the hopfion profile, causing a breakdown of the solitonic configuration. Formation of mazelike hopfions was also observed in Ref. \cite{Hu2025}.

Among the isolated hopfions, those with $Q_H = +1$ consistently exhibit the highest eigen-energy (Fig.~\ref{fig04}(f)). In contrast, the eigen-energies of hopfions with $Q_H = 0$ and $Q_H = -1$ display a crossover behavior. This energetic ordering correlates with the internal structure and topological character of each configuration: $Q_H = +1$ hopfions incur a higher energy cost due to their fully linked preimage topology, while the lower energy of $Q_H = -1$ hopfions results from the compactness of their constituent bimeron textures, whose crescent-shaped parts face inward, thereby minimizing distortions in the surrounding homogeneous field.

\begin{figure*}[t]
  \centering
  \includegraphics[width=0.8\linewidth]{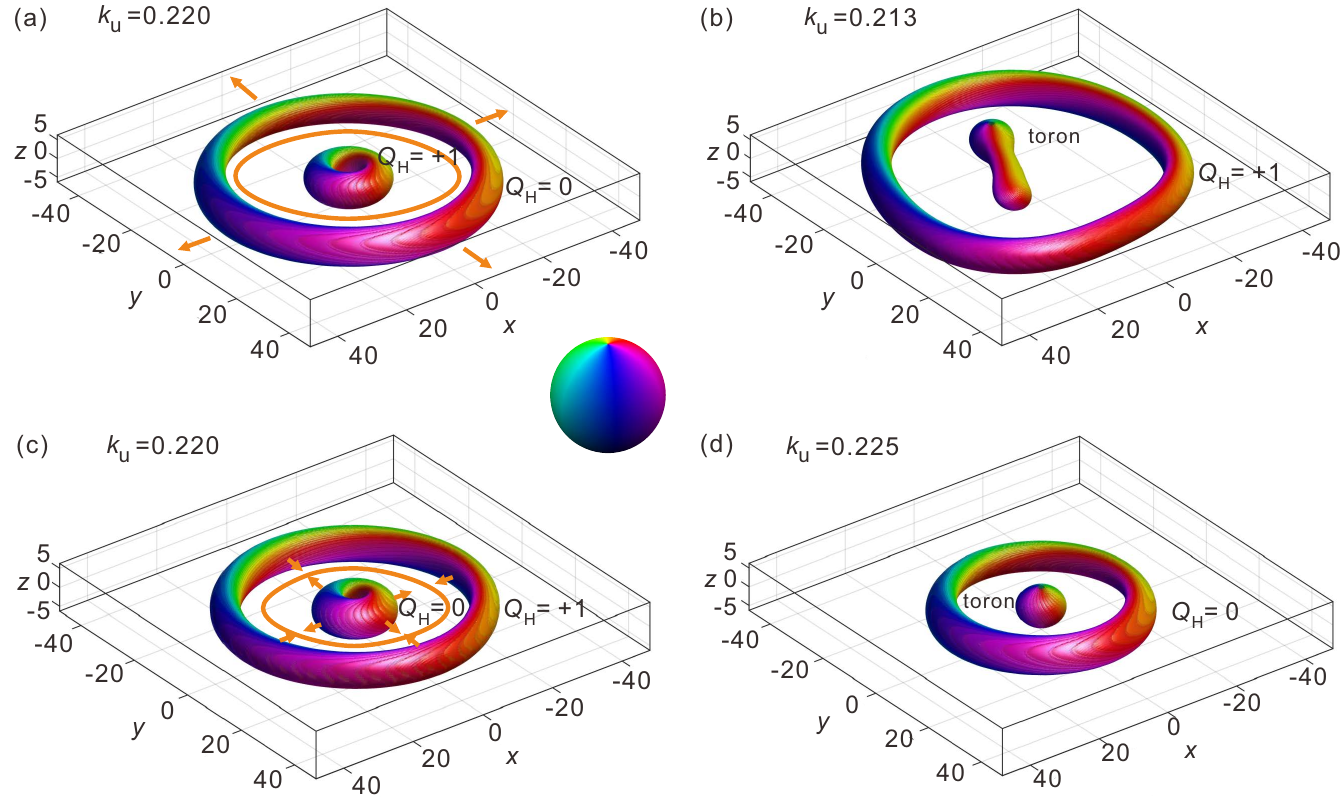}
  \caption{\label{fig05} \textrm{Hopfion bags.} 
(a, b) Hopfion bag configurations formed by outer hopfions with vanishing eigen-energies, allowing their radii to flexibly adjust and minimize overlap with the enclosed structures. In (a), the interior is occupied by another hopfion, while in (b) it consists of an elongating toron, which nucleates a CF-1 finger phase. 
(c, d) Hopfion bags stabilized as a compromise between overlapping hopfions with non-equilibrium radii. In (d), the inner hopfion collapses into a toron due to excessive overlap. }
\end{figure*}

\section{Hopfion bags.}

The dependence $R_0(k_u)$ in Fig.~\ref{fig04}(c) also indicates the possibility of assembling composite hopfions (Fig.~\ref{fig05}), consistent with an effective repulsive interaction between individual hopfions. Before analyzing such composite structures, we first discuss the nature of the repulsion between isolated hopfions.

\subsection{Repulsive interactions between isolated hopfions \label{A4}}

\begin{figure}[t]
    \centering
    \includegraphics[width=0.48\textwidth]{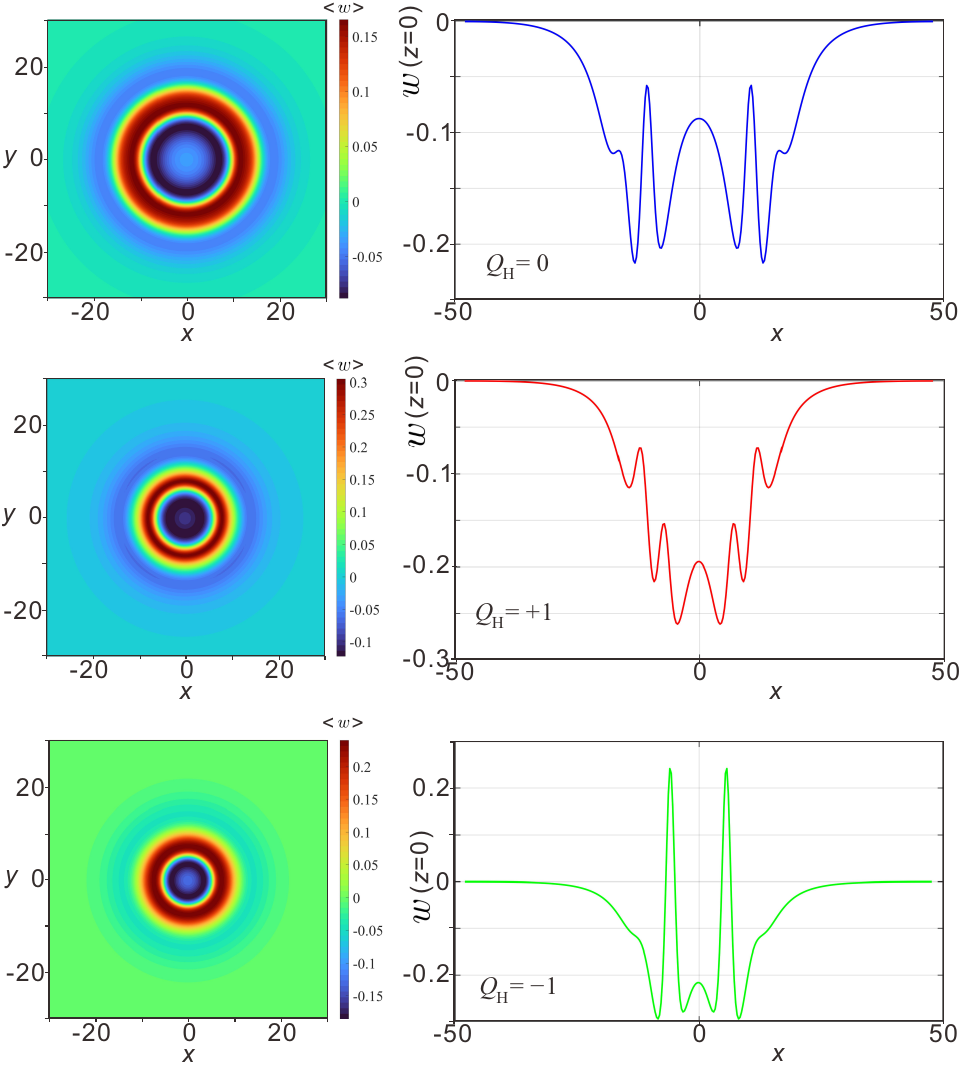}
    \caption{\textrm{Energetic fingerprints of isolated hopfions.} 
    The energetic characteristics of isolated hopfions are presented as color plots of the energy density on the $xy$-plane, averaged over the film thickness, together with the corresponding energy profiles along the middle cross-section of the film.
    }
    \label{fig05v}
\end{figure}

The repulsive inter-hopfion interaction originates from the characteristic energy density asymptotics within hopfion structures (Fig.~\ref{fig05v}). Both the energy distribution averaged over the film thickness and the local energy density in individual cross-sections exhibit negative values at the hopfion outskirts. This implies that any spatial overlap of the energy profiles from neighboring hopfions results in an energy increase, thereby generating an effective repulsive interaction.

The inter–hopfion interaction potential can be quantified by preparing two hopfion configurations at controlled separations and computing the total system energy as a function of their distance. Direct simulations of this type, however, are computationally demanding, as they require large simulation domains and fine spatial resolution to avoid boundary-induced artifacts. As an efficient alternative, the interaction potential can be estimated from the asymptotic decay of the energy–density profiles associated with an isolated hopfion. By analyzing the overlap of these tails, one can infer the effective interaction strength without performing full two–hopfion simulations.

\begin{figure*}[h!]
    \centering
    \includegraphics[width=0.8\textwidth]{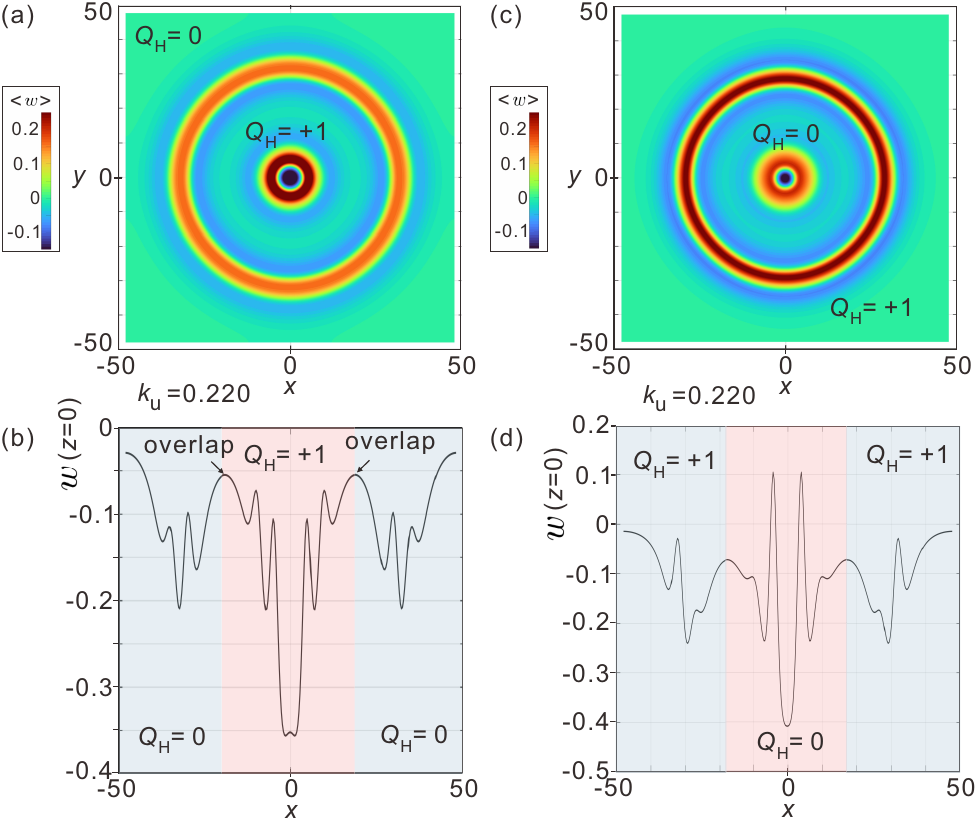}
    \caption{\textrm{Energetic fingerprints of composite hopfions.} 
    The energetic characteristics of composite hopfions are presented as color plots of the energy density on the $xy$-plane, averaged over the film thickness, together with the corresponding energy profiles along the middle cross-section of the film.
    }
    \label{fig06v}
\end{figure*}

Composite hopfions (Fig.~\ref{fig06v}) provide a complementary manifestation of this repulsive interaction, where the formation of nested multi-soliton configurations necessarily involves energy penalties due to profile overlap and deviations from the equilibrium radii of the constituent hopfions.

\subsection{Composite hopfions}
The most natural configuration within a composite hopfion involves placing a $Q_H = \pm 1$ hopfion inside the interior of an inflating $Q_H = 0$ hopfion (Fig.~\ref{fig05}(a)). Since the eigen-energy of the $Q_H = 0$ hopfion remains relatively close to that of the homogeneous state—vanishing as $k_u$ approaches its critical value—it preferentially increases its radius to minimize inter-hopfion repulsion, while the internal $Q_H = \pm 1$ hopfion largely retains its structure.

For $k_u = 0.22$, the eigen-energies of isolated hopfions are $W(Q_H = 0) = 42.06$ and $W(Q_H = 1) = 86.45$ (Fig.~\ref{fig04}(f)). The total energy of the corresponding composite hopfion configuration is $W = 168.04$. This value allows us to estimate the excess energy $\Delta W = 39.53$, arising from the mismatch of the hopfion radii relative to their equilibrium sizes and from the inter-soliton interactions within the composite structure. The dominant contribution to this excess energy originates from the overlap of the magnetization profiles in the region between the nested solitons. 

When this excess energy exceeds the energy barrier separating the internal hopfion state from a toron configuration, a collapse of the inner hopfion can occur, resulting in the formation of a composite texture where a $Q_H = 0$ hopfion encloses a toron (Fig. \ref{fig05} (d)). The same mechanism can be extended to encompass multiple torons within a $Q_H = 0$ hopfion, which, being near its inflation point and possessing minimal eigen-energy, imposes the least pressure on its interior. 

The opposite ordering of hopfions, with a $Q_H=0$ hopfion placed inside a $Q_H=+1$ hopfion, results in a significantly more stressed composite configuration with a total energy $W = 223.82$ (Fig.~\ref{fig05}(c)). In this arrangement, both hopfions possess radii markedly displaced from their respective equilibrium values: the outer $Q_H=+1$ hopfion tends to shrink, while the inner $Q_H=0$ hopfion seeks to expand. This mutual constraint increases their spatial overlap, leading to a substantial excess energy $\Delta W = 95.31$, primarily originating from the distorted profiles and enhanced inter-soliton interaction. 

Interestingly, one can envision yet another bag-type configuration realized between the critical anisotropy values for the two hopfion types, for example at $k_u=0.213$ (Fig.~\ref{fig05}(b)), where the outer $Q_H = +1$ hopfion inflates without energy cost, while the interior accommodates the CF-1 finger phase. Notably, this modulated interior phase can be nucleated through the elongation of a toron.

The proposed bag configurations extend the conventional concept of a bag, which typically refers to a bundle of skyrmions enclosed by a circular domain wall~\cite{foster2019,Tang2021}.


\begin{figure*}[t]
  \centering
  \includegraphics[width=0.8\linewidth]{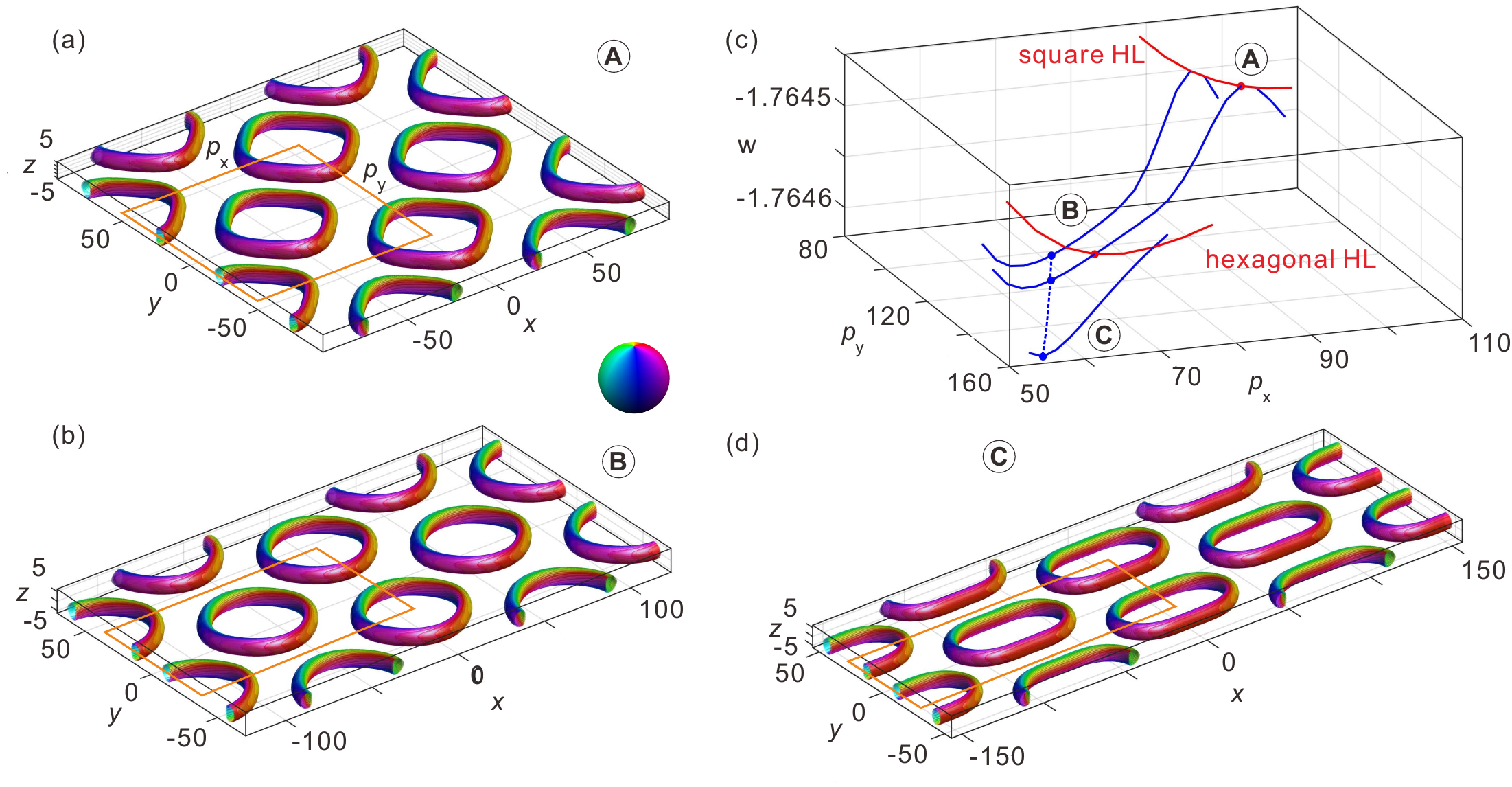}
  \caption{\label{fig06} \textbf{Instability of hopfion lattices.} (a, b) Preimage structures of square and hexagonal HLs, each corresponding to energy minima when the unit cell geometry is fixed. (c) Energy densities of square and hexagonal HLs, plotted as red curves with red dots marking energy minima. Blue curves show the energy variation with changing unit cell size, revealing the instability of HLs under lattice deformation. (d) Elongated HL structure attempting to reach the global energy minimum by transitioning toward the corresponding modulated finger phase.}
\end{figure*}

\section{Unstable hopfion lattices: elongation into finger phases}

Another important conclusion emerging from the $R_0(k_u)$ behavior in Fig.~\ref{fig04}(c) is the fundamental instability of hopfion lattices in this system. In particular, the eigen-energy of an expanding hopfion never becomes negative and therefore does not trigger condensation into a periodic modulated phase. This is in stark contrast to spiral and skyrmion states, where isolated solitons retain a finite equilibrium size, and the point at which their eigen-energy crosses zero marks the onset of lattice formation~\cite{Togawa12}. Hence, isolated hopfions cannot serve as stable building blocks for periodic hopfion lattices in the present confined geometry. Any attempt to realize a hopfion lattice must therefore occur within the intrinsic stability region of the corresponding finger phases.

To explore this scenario, we computed the total energy of a $Q_H=1$ hopfion lattice while systematically varying the unit-cell dimensions $p_x$ and $p_y$ (the orange rectangles in Fig.~\ref{fig06}). For a square unit cell (Fig.~\ref{fig06}(a)), the energy as a function of side length appears to exhibit a minimum along this one-dimensional cut of the parameter space (red curve in Fig.~\ref{fig06}(c)). A comparable behavior is found for a hexagonal cell (Fig.~\ref{fig06}(b)), where the ratio $p_y=\sqrt{3}p_x$ is held fixed. However, when the full two-dimensional space of $(p_x,p_y)$ is explored, it becomes clear that these apparent minima are not true minima of the total energy: the square-cell minimum corresponds to a saddle point, while the minimum along the $\sqrt{3}$ ratio lies on a descending slope of the actual three-dimensional energy landscape. 

The genuine global minimum is obtained only when the unit cell is elongated, forcing the hopfions themselves to stretch and deform (Fig.~\ref{fig06}(d)). This elongation ultimately transforms the nominal hopfion lattice into a modulated finger phase whose period is fixed by the system parameters. This behavior highlights a key distinction between skyrmion-based and hopfion-based systems: the inherently three-dimensional topology of hopfions prevents the formation of stable periodic lattices under confinement, as any periodic arrangement inevitably relaxes into the lower-energy finger state.

Interestingly, a hexagonal hopfion lattice was theoretically reported in Ref.~\cite{Tai} under applied magnetic fields. However, such results must be examined over the full range of unit-cell parameters $(p_x,p_y)$ to confirm true stability. As shown above, the apparent minimum obtained by fixing $p_y=\sqrt{3}p_x$ is not a true minimum, but rather a point located on a slope of the full energy surface.

The same phenomenon of meta-matter instability was recently reported for skyrmioniums in quasi-2D chiral magnets \cite{Nakamura}, where both hexagonal and square skyrmionium lattices were shown to elongate and eventually transform into spiral states. This parallel is not accidental: skyrmioniums represent the natural two-dimensional cross-sections of hopfions, and thus their behavior encodes important aspects of the underlying three-dimensional topology. The observed lattice instability therefore reflects a more general and universal mechanism.

In both cases—skyrmioniums in thin films and hopfions under confinement—the constituent solitons possess internal degrees of freedom that make them susceptible to elongation rather than isotropic packing. As a result, periodic arrays of such solitons do not form true equilibrium states; instead, they continuously deform into the corresponding lower-energy modulated phases (spirals in 2D, finger phases in 3D). This universality highlights an important organizing principle: when the soliton’s internal structure is compatible with a continuously deformable modulated state, lattice formation becomes unstable, and the system naturally relaxes toward an extended periodic texture.

Thus, the instability of hopfion lattices can be understood as the three-dimensional analogue of skyrmionium-lattice instability: in both cases, the solitons are not rigid entities but can stretch along a preferred direction, enabling a smooth transformation into the energetically favorable modulated background. This analogy further strengthens the interpretation of hopfions as 3D counterparts of skyrmioniums and points to a unifying framework for understanding metastable “topological meta-matter’’ across different dimensionalities and material platforms.

\section{Discussion and conclusions}
In this work, we have systematically explored the stabilization mechanisms, structural diversity, and interactions of three-dimensional hopfions in confined chiral magnetic films with surface anchoring. Our results reveal that the metastability regions of distinct hopfion types are closely linked to the phase boundaries of the underlying modulated finger textures. As the uniaxial anisotropy decreases (corresponding to a reduced electric field in liquid crystalline analogues), isolated hopfions inflate and progressively deform, eventually transforming into extended modulated phases: CF-1 phases for $Q_H = 0$ hopfions and CF-2 phases for $Q_H = \pm 1$ types. 

This behavior exemplifies a broader principle whereby isolated solitons embedded within a homogeneous ferromagnetic background act as sensitive probes of nearby modulated phases in the topological phase diagram.
%
Torons, when prevented from assembling into periodic lattices, likewise elongate and fill the available space, reproducing the CF-1 phase (Fig. \ref{fig05} (b)). Similarly, defect-free isolated skyrmions stabilized at low surface anchoring strengths destabilize through elliptical instabilities, initiating the formation of spiral states~\cite{Leonov2016}. Notably, composite hopfions in such parameter regimes generate mixed CF-1 and CF-2 phases, with their equilibrium periods determined by the energy minima of the corresponding modulated textures.

A key conclusion of this study is the intrinsic instability of periodic hopfion lattices under the confinement conditions considered here. Whereas lattices of two-dimensional solitons, such as skyrmions and torons, are stabilized by mutual interactions and confinement effects, hopfion lattices lack a local energy minimum and spontaneously relax into modulated finger phases. However, this limitation could potentially be overcome by modifying the energetic hierarchy of competing modulated phases. For example, if the conical phase were stabilized as the global energy minimum, hopfions could exist as localized inclusions within a conical background. In such a setting, hopfions would experience an effective attraction mediated by the surrounding conical phase, enabling the formation of stable clusters or even periodic hopfion lattices embedded in a conical matrix.

Our findings thus clarify the delicate interplay of energetic and topological constraints governing the stability of three-dimensional solitons in confined chiral systems and suggest practical routes for topological phase control via external fields, elastic tuning, or confinement engineering. Moreover, the demonstrated capacity to stabilize multi-soliton configurations through hierarchical nesting introduces a versatile strategy for constructing tunable, metastable textures with potentially reconfigurable properties. Future work should address the dynamical stability, nucleation pathways, and inter-soliton interactions in time-dependent or nonequilibrium conditions, opening new opportunities for topological spintronic and photonic devices based on three-dimensional solitonic matter.

\textbf{Acknowledgements.} The authors would like to thank Paul Sutcliffe, Asha Ghanghas and Jun-Yong Lee for interesting discussions.


\end{document}